\documentclass[pre,11pt,preprint,superscriptaddress, amsmath, amssymb, floatfix]{revtex4-1} 

\usepackage{graphicx}  
\usepackage{graphics}
\usepackage{dcolumn}   
\usepackage{bm}        
\usepackage{amssymb}   
\usepackage{color}
\usepackage{amsmath}

\newcommand{\la}{\langle}
\newcommand{\ra}{\rangle}

\begin{document}



\title{Emergence of Bimodality in Controlling Complex Networks}
\author{Tao Jia}
\affiliation{Center for Complex Network Research and Department of
  Physics,Northeastern University, Boston, Massachusetts 02115, USA} 
\author{Yang-Yu Liu}  
\affiliation{Center for Complex Network Research and Department of
  Physics,Northeastern University, Boston, Massachusetts 02115, USA} 
\affiliation{Center for Cancer Systems Biology, Dana-Farber Cancer Institute, Boston, Massachusetts 02115, USA}
\author{Endre Cs\'oka}
\affiliation{E\"otv\"os Lor\'and University, H-1053 Budapest, Hungary}
\author{M\'arton P\'osfai}
\affiliation{Center for Complex Network Research and Department of
  Physics,Northeastern University, Boston, Massachusetts 02115, USA} 
\affiliation{Department of Theoretical Physics, Budapest
   University of Technology and Economics, H-1521 Budapest, Hungary}
\affiliation{Department of Physics of Complex Systems, E\"otv\"os
  Lor\'and University, H-1053 Budapest, Hungary}
\author{Jean-Jacques  Slotine}
\affiliation{Nonlinear Systems Laboratory, Massachusetts Institute of Technology, Cambridge, Massachusetts 02139, USA}
\affiliation{Department of Mechanical Engineering and Department of Brain and
Cognitive Sciences, Massachusetts Institute of Technology, Cambridge, Massachusetts 02139, USA} 
\author{Albert-L\'{a}szl\'{o} Barab\'{a}si}
\email{barabasi@gmail.com}
\affiliation{Center for Complex Network Research and Department of
Physics,Northeastern University, Boston, Massachusetts 02115, USA} 
\affiliation{Center for Cancer Systems Biology, Dana-Farber Cancer Institute, Boston, Massachusetts 02115, USA}
\affiliation{Department of Medicine and Division of Network Medicine, Brigham and Women's Hospital, Harvard Medical School, Boston, Massachusetts 02115, USA} 


\begin{abstract}
Our ability to control complex systems is a fundamental challenge of contemporary science. Recently introduced tools to identify the driver nodes, nodes through which we can achieve full control, predict the existence of multiple control configurations, prompting us to classify each node in a network based on their role in control. Accordingly a node is critical, intermittent or redundant if it acts as a driver node in all, some or none of the control configurations. Here we develop an analytical framework to identify the category of each node, leading to the discovery of two distinct control modes in complex systems: centralized vs distributed control. We predict the control mode for an arbitrary network and show that one can alter it through small structural perturbations. The uncovered bimodality has implications from network security to organizational research and offers new insights into the dynamics and control of complex systems.

\end{abstract}

\maketitle 

\begin{figure}[ht]
\begin{center}
\includegraphics{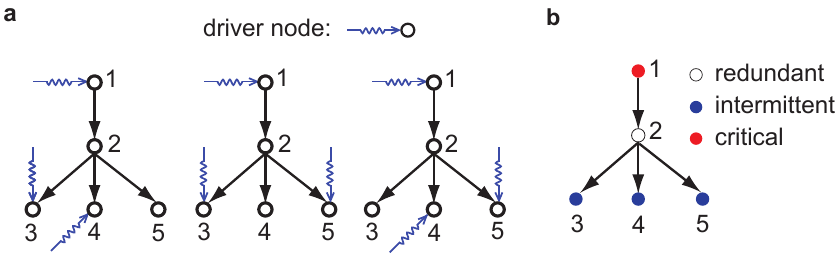}
\caption{
{\bf (a)} A network with five nodes that can be controlled via three driver nodes ($N_\text{D} = 3$). The system is characterized by three distinct minimum driver node sets (MDS's).
{\bf (b)} Node 1 is critical as it is part of all MDS's shown in (a), node 2 is redundant as it does not participate in any MDS and nodes 3, 4, 5 are intermittent, participating in some but not all MDS's. 
\label{fig:fig1}}
\end{center}
\end{figure}

A dynamical system is controllable if it can be driven from any initial state to any desired final state within finite time \cite{Kalman-JSIAM-63,Luenberger-Book-79,Chui-Book-89,Slotine-Book-91}. In general, controllability can be achieved by changing the state of a small set of driver nodes, which drive the dynamics of the whole network \cite{Liu-Nature-11}. For a linear time-invariant dynamics we can identify the minimum driver node set (MDS), representing the smallest set of nodes through which we can yield control over the whole system \cite{Liu-Nature-11, Rajapakse-pnas-2011, Lestas-nature-2010, Nepusz-NP-12,Yan-PRL-12,Liu-PO-12,Wang-PRE-12, tang-plosone-2012, cowan-plosone-2012, gutierrez-sr-2012, wang2012control, mones2012hierarchy, pu2012robustness, Posfai2013, Liu-pnas-2013}. While the number of driver nodes sufficient and necessary for control ($N_{\text{D}}$) is primarily fixed by the network's degree distribution, there are multiple MDS's with the same $N_\text{D}$ that can maintain control. For example, for the five-node network shown in Fig. \ref{fig:fig1}a $N_{\text{D}} = 3$, but the formalism indicates that control can be achieved via three different MDS's: $\{1,3,4\}$, $\{1,3,5\}$ and $\{1,4,5\}$.

Here we explore the role of individual nodes in controlling a network by classifying each node into one of three categories based on its likelihood of being included in MDS \cite{Commault-IJACSP-11}: critical, meaning that a node must always be controlled in order to control a system (it is part of all MDS's); redundant, meaning that it is never required for control (does not participate in any MDS's) and intermittent, meaning that it acts as driver node in some control configurations, but not in others. For example, in Fig. \ref{fig:fig1}a node 1 is critical, node 2 is redundant and nodes 3, 4, 5 are intermittent (Fig. \ref{fig:fig1}b). This classification leads to the discovery of a bifurcation phenomenon, predicting that a bimodal behavior determines the controllability of many real networks. This bimodality helps us uncover two control modes, centralized vs distributed. We demonstrate both analytically and numerically the existence of these two modes and show that the predicted control modes naturally emerge in a wide range of real networks.

{\bf Results} 

{\bf Identifying node categories.}
We developed an algorithm to identify the redundant nodes in a network with $N$ nodes and $L$ links in $\mathcal O(NL)$ steps, offering the fraction $n_\text{r}$ of redundant nodes for an arbitrary network (see Methods). We also proved that a node is critical if and only if it has no incoming links (Supplementary Note 1), a theorem that provides the fraction of critical nodes in a network as $n_\text{c} = P_{\text{in}}(0)$ where $P_{\text{in}}(k)$ is the incoming degree distribution. Finally intermittent nodes are neither critical nor redundant, hence their fraction is $n_\text{i} = 1 - n_\text{c} - n_\text{r}$.

\begin{figure}[ht]
\begin{center}
\includegraphics{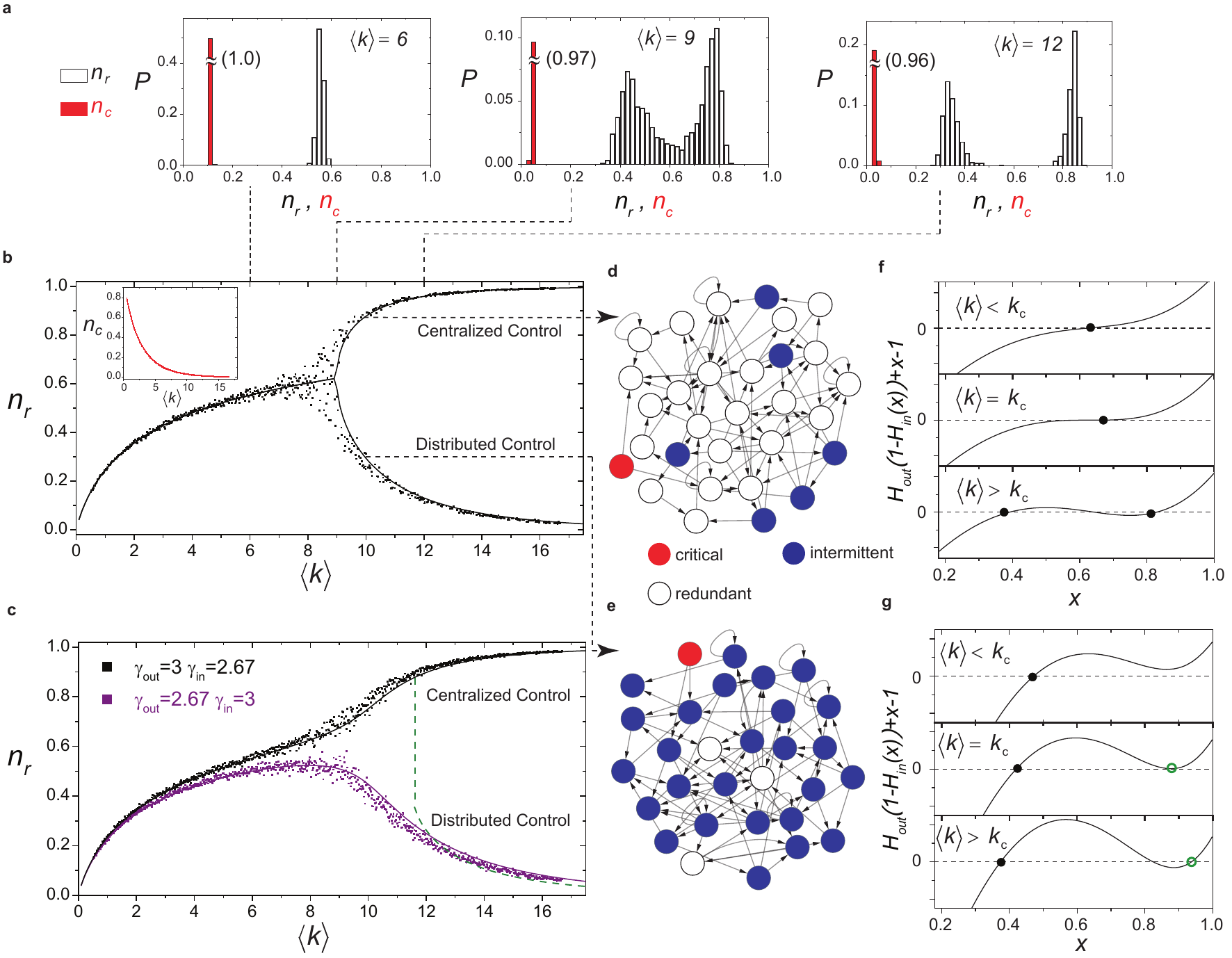}
\caption{
 {\bf Bifurcation in control.} 
{\bf (a)} Distribution of the fraction of redundant ($n_\text{r}$) and critical ($n_\text{c}$) nodes in scale-free networks with $\gamma_{\text{out}} = \gamma_{\text{in}} =3$, documenting the emergence of a bimodal behavior for high $\la k \ra$. 
{\bf (b)} $n_\text{r}$ and $n_\text{c}$ (insert) $vs$ $\la k \ra$ in scale-free networks with degree exponents $\gamma_{\text{out}} = \gamma_{\text{in}} =3$, illustrating the emergence of a bifurcation for high $\la k \ra$. 
{\bf (c)} $n_\text{r}$ in scale-free networks with asymmetric in- and out-degree distribution, $i.e.$ $\gamma_{\text{out}} = 3$, $\gamma_{\text{in}} = 2.67$ (upper branch) and $\gamma_{\text{out}} = 2.67$, $\gamma_{\text{in}} = 3$ (lower branch). The control mode is pre-determined by their degree asymmetry. The solid lines in (b) and (c) correspond to the analytical prediction of equation (\ref{eq:beta_equation}). The dashed line is the discontinuous solution of equation (\ref{eq:beta_equation}) when $\gamma_{\text{out}} = 3$ and $\gamma_{\text{in}} = 2.67$, which shows a gap between the actual evolution of $n_\text{r}$. 
{\bf (d,e)} Networks displaying centralized and distributed control. For both networks $N_\text{D} = 4$ and $N_\text{c} = 1$, yet they have rather different number of redundant nodes, $N_\text{r} = 23$ in (d) and $N_\text{r} = 3$ in (e).
{\bf (f,g)} The curves $H_{\text{out}}\bigl(1- H_{\text{in}}(x) \bigr)+x -1$ from equation (\ref{eq:beta_equation}) shown for different $\la k \ra$, $\gamma_{\text{out}} = \gamma_{\text{in}} =3$ in (f) and $\gamma_{\text{out}} = 3$, $\gamma_{\text{in}} = 2.67$ in (g). The continuous stable solutions of equation (\ref{eq:beta_equation}) are shown as filled dots and the discontinuous solutions as empty circles. 
\label{fig:nr_no_nc}}
\end{center}
\end{figure}

{\bf Bimodality in control.}
To explore the role of the network topology we measured $n_\text{r}$ and $n_\text{c}$ for networks with varying average degree $\la k \ra$. We find that for small $\la k \ra$ for an ensemble of networks with identical degree distribution $P(k)$, $n_\text{r}$ and $n_\text{c}$ follow narrow distributions (Fig. \ref{fig:nr_no_nc}a). This means that $n_\text{r}$ and $n_\text{c}$ are primarily determined by $P(k)$~\cite{Molloy-RSA-95, Newman-PRE-01, Albert-RMP-02, Kim-NJP-12}. Surprisingly, when $\la k \ra$ exceeds a critical value $k_c$~\cite{Liu-PRL-12}, $P(n_\text{r})$ becomes bimodal, implying that systems with the same $P(k)$ can exist in two distinct states: some have small $n_\text{r}$ and for others $n_\text{r}$ is very large. This bimodality is present in both random~\cite{Erdos-PMIHAS-60} and scale free networks~\cite{Barabasi-Science-99,Caldarelli-Book-07} (Fig. \ref{fig:nr_no_nc}c, Supplementary Note 2). The emergence of this bimodal behavior is best captured by plotting $n_\text{r}$ $vs$ $\la k \ra$, observing a bifurcation as $\la k \ra$ reaches $k_c$ (Fig. \ref{fig:nr_no_nc}b). This bifurcation predicts two distinct control modes:

Centralized control:
For networks that follow the upper branch of the bifurcation diagram most nodes are redundant ($n_\text{r}$). This means that in these networks one can achieve control through a small fraction of all nodes ($n_\text{c} + n_\text{i}$), hence capturing a centralized control mode (Fig. \ref{fig:nr_no_nc}d). This has obvious consequences in communication systems, as one can ensure a centralized system's security by protecting only a small fraction of nodes ($n_\text{c} + n_\text{i}$); in an organizational setting centralized network may be better suited for task execution, like manufacturing, where efficiency is enhanced by subordination to a small number of control nodes.

Distributed control:
For networks on the lower branch $n_\text{c} + n_\text{i}$ can exceed 90\% of the nodes. Hence most nodes can act as driver nodes in some MDS's, resulting in a distributed control mode (Fig. \ref{fig:nr_no_nc}e). Securing such distributed communication networks requires significant resources, as one can gain control of the system via a large number of control configurations. Yet organizations displaying distributed control may be more capable of harboring innovation, as different node combinations could take control of the organization's direction.

Intuitively, one would expect these two different control modes to be associated with distinct network properties. However, we find that networks with identical in- and out-degree distribution can develop centralized or distributed control modes with equal probability (Fig. \ref{fig:nr_no_nc}a). For networks with different in- and out-degree distributions the symmetry between the two branches is broken, forcing the network in one or the other branch of the bifurcation diagram. This is illustrated for networks with $P(k_\text{in}) \sim k_\text{in}^{-\gamma_\text{in}}$ and $P(k_\text{out}) \sim k_\text{out}^{-\gamma_\text{out}}$ where $\gamma_\text{in} \neq \gamma_\text{out}$. The degree asymmetry forces a network to follow one or the other branch of the bifurcation diagram (Fig. \ref{fig:nr_no_nc}c), pre-determining whether the network displays a centralized or a distributed control mode.

{\bf Analytical approach.} 
To understand the origin of the observed control modes, we map the control problem to maximum matching \cite{Liu-Nature-11,Hopcroft-SIAM-73}, which provides the fraction of redundant nodes in infinite networks as
\begin{equation}
n_\text{r}  = 1-G_{\text{in}}(\theta_{\text{out}})
\label{eq:nr_analytical},
\end{equation}
where $G_{\text{in}}(x) = \sum_{k=0}^{\infty} x^k P_\text{in}(k)$ is the
generating function for $P_{\text{in}}(k)$~\cite{Newman-PRE-01}.

Here $\theta_{\text{out}}$ is the solution of the recursive equation
\begin{equation}
1-\theta_{\text{out}} = H_{\text{out}}\bigl(1- H_{\text{in}}(\theta_{\text{out}}) \bigr), \label{eq:beta_equation}
\end{equation}
where $H_{\text{out, in}}(x) = \sum_{k=1}^{\infty}x^{k-1}Q_{\text{out,in}}(k)$
and $Q_{\text{out,in}}(k) = P_{\text{out,in}} (k)\times k/\la k \ra$ is the
excess degree distribution. Equations (\ref{eq:nr_analytical}) and (\ref{eq:beta_equation}) self-consistently predict $n_\text{r}$ from the network's degree distribution, in excellent agreement with the numerical results (see the continuous lines in Fig. \ref{fig:nr_no_nc}b,c), helping us understand the origin of the observed bifurcation. For $P_{\text{in}} (k) = P_{\text{out}} (k)$ equation (\ref{eq:beta_equation}) has a single solution for small $\la k \ra$ (top panel in Fig. \ref{fig:nr_no_nc}f). As $\la k \ra$ reaches $k_c$, equation (\ref{eq:beta_equation}) develops three solutions, two of which are stable, corresponding to the two branches of the bifurcation diagram. As the new solutions are an analytic continuation of the $\la k \ra < k_c$ solution, the system can continue in any of the two branches, hence the centralized or the distributed control modes are equiprobable.  If, however, $P_{\text{in}} (k) \ne P_{\text{out}} (k)$, for $\la k \ra > k_c$ only one of the two stable solutions is an analytical continuation of the pre-bifurcation solution (middle panel in Fig. \ref{fig:nr_no_nc}g). In other words infinite systems with different $P_\text{out}(k)$ and $P_\text{in}(k)$ are destined for the centralized or distributed control mode, corresponding to the two branches of the bifurcation diagram, depending on the nature of their degree asymmetry. For small systems, the two modes can coexist and jumps between them are possible (Supplementary Figs. S1 and S2).

\begin{figure}[ht]
\begin{center}
\includegraphics{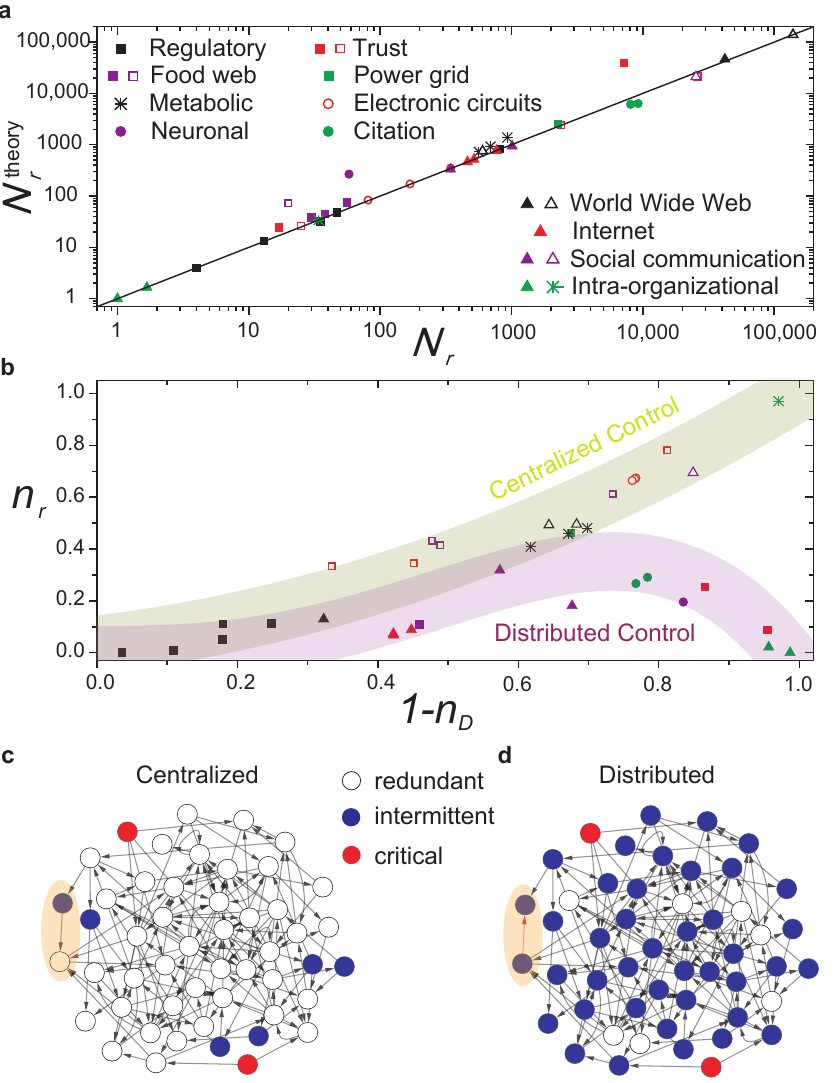}
\caption{
{\bf Bimodality in real networks.} {\bf (a)} The number of redundant nodes $N_\text{r}$ in real networks (Supplementary Tables S1 and S2) and the theoretical prediction of equations (\ref{eq:nr_analytical}) and (\ref{eq:beta_equation}). Hollow symbols represent networks characterized by centralized control ($\Delta n_\text{r} > 0$) and solid symbols correspond to the distributed control mode ($\Delta n_\text{r} < 0$). For networks marked with a star symbol, $\Delta n_\text{r} \approx 0$ and the control mode can not be identified. 
{\bf (b)} $n_\text{r}$ vs $1-n_\text{D}$ for real networks. The networks follow the two branches corresponding to centralized and distributed control (highlighted bands).
{\bf (c)} A network in the centralized control mode. 
{\bf (d)} After flipping the direction of one link in (c) (highlighted in both (c) and (d)), the control configuration changes from centralized to distributed mode. For both networks $N_\text{D} = 5$ and $N_\text{c} = 2$, but in the centralized mode 8 nodes play a role in control (2 critical plus 6 intermittent) while in the distributed mode 43 nodes are engaged in control (2 critical plus 41 intermittent).
\label{fg:real_network}}
\end{center}
\end{figure}

{\bf Bimodality in Real Networks.} 
To demonstrate the empirical relevance of these tools, we used equations (\ref{eq:nr_analytical}) and (\ref{eq:beta_equation}) to calculate $n_\text{r}$ for several real networks, starting from their degree distribution. Note that while for some of these networks control is of potential relevance (like regulatory networks \cite{Milo-Science-02, Babu-JMB-06, Collado-Vides-NAR-08} or neural networks \cite{White1986}), others like citation networks~\cite{Redner1998, Leskovec-05, Redner2005} or the WWW~\cite{Albert-Nature-99,Adamic-05,Leskovec-arXiv-08} are of little or no relevance for control. We analyze them mainly because they offer diverse topologies that test the limits of our predictions.
We obtain a reasonable agreement between the analytically predicted $N^{\text{theory}}_\text{r}$ and $N_\text{r}$ obtained directly for each network (Fig. \ref{fg:real_network}a, Supplementary Fig. S3). 
To compare $n_\text{r}$ for different networks, we plot them in function of $1-n_\text{D}$, representing the fraction of nodes that do not require external control. Obviously, real networks have widely different degree distributions, degree asymmetries and even potential correlations~\cite{Ravasz-science-2002, Newman-PRL-02,Pastor-Satorras-PRL-01b,Maslov-Science-02,Lee-EPJB-06,Posfai2013}.
Despite these the predicted bifurcation is observed in real systems as well (Fig. 2b, Supplementary Figure S4). We find that some network structures, like the {\it E. coli} metabolic network \cite{Jeong-Nature-00}, are in the ``pre-bifurcation" region $\la k \ra < k_c$; others, like the mobile call network \cite{cesar-physca-2006, Onnela2007, Gonzalez-nature-2008, Song-Science-10, Jiang-pnas-2013} and citation networks \cite{Redner1998, Leskovec-05, Redner2005}, follow one or the other branch of the bifurcation diagram, indicating that they are characterized by either centralized or distributed control. 
To identify the control mode charactering a particular network, we introduce the transpose network, whose wiring diagram is identical to the original network but the direction of each link is reversed. The control mode is captured by comparing $n_\text{r}$ of a network with $n_\text{r}^\text{T}$ of its transpose network: if $\Delta n_\text{r} = n_\text{r} - n^\text{T}_\text{r} > 0$ a network is centralized and if $\Delta n_\text{r} < 0$ it is distributed (Supplementary Note 3). We find that a network's degree distribution allows us to infer its control mode (Supplementary Note 4).

{\bf Altering the Control Mode.} 
The fact that the two control modes are better suited for different tasks raises an important question: can we turn a network initially in the centralized mode into a network in the distributed mode, or the other way around? Such a transition is achieved by the transpose network. Indeed, we can show that if the original network has a distributed control mode, its transpose will be centralized. Yet, in most real systems switching the direction of all links is either impossible or infeasible. We find, however, that in some networks the transition can be induced by local changes, in some cases requiring us to flip the direction of only a single well chosen link (Fig. \ref{fg:real_network}c). The precise local change required to induce the transition is determined by the
network's size and degree asymmetry: the larger the network or the higher degree asymmetry, the more links need to be altered to induce this transition. For example, for two manufacturing and consulting networks \cite{Parker-Book-04} ($N=77$, $L=2228$ and $N=46$, $L=858$) the control mode can be altered by flipping only one link. In contrast for the prison inmate network \cite{Duijn-JMS-03,Alon-Science-04}, food web in Little Rock lake \cite{Martinez-EM-91} and the {\it C. elegans} neural network \cite{White1986} we need to change the direction of several links (4, 17 and 54 respectively, representing 2\%, 0.7\% and 2\% of all links) to alter the control mode.

{\bf Discussion}

In summary, our attempts to quantify the role of individual nodes have led us to an unexpected discovery: the existence of two distinct control modes, whose emergence is governed by a bifurcation phenomenon. The control mode is determined by the
system's degree asymmetry and has immediate implications on network security or efficiency in organizational settings. Most importantly, bimodality is a general property of all dense networks, hence we must consider its implications each time we wish to control a complex system with sufficient connectivity.

These results raise several intriguing questions. For example, the finding that the control mode can be altered via structural perturbations raises the need for tools to identify the minimum number of links whose reversal can help us reach the desired control mode \cite{chenguangrong2000}. Furthermore, the proposed node classification raises the possibility to correlate the role of a node in control with the intrinsic node attributes. Finally, our node classification is based on the node's participation in various MDS's. However, one can explore other node classifications as well, like that based on the energy needed for control, or the time necessary to reach the final state \cite{Yan-PRL-12}, important issues that need further investigation.
While bimodality is ubiquitous in physics, chemistry and biology, the finding that it also plays a key role in network control opens new avenues to explore the control of real systems. 

\newpage

{\bf Methods}

{\bf Identifying the driver nodes.} We convert a directed network into a bipartite graph with two disjoint sets of $out$ and $in$ nodes. A directed link from node $i$ to $j$ corresponds to a connection between node $i$ in the $out$ set and node $j^{\prime}$ in the $in$ set (Supplementary Fig. S5). By finding the maximum matching of the bipartite graph, the minimum driver nodes are unmatched nodes in the $in$ set. The critical, intermittent and redundant nodes are always matched, occasionally matched and never matched nodes in the $in$ set, respectively (Supplementary Note 5).

{\bf Identifying the redundant nodes.} Redundant nodes are always matched in the bipartite graph. Therefore if we force them unmatched, there would be no alternative matching and the number of matched nodes will decrease, inspiring the algorithm we developed to identify them:

(1) Find the maximum matching of the bipartite graph using the Hopcroft-Karp algorithm \cite{Dion-Automatica-03,Zdeborova-JSM-06,Karp-IEEE-81,Hopcroft-SIAM-73,Lovasz-Book-09}. Obtain a set of matched nodes in the $in$ set (denoted by $M$).
(2) Pick one element (denoted by node $i$) in $M$. Identify the node in the $out$ set that matches node $i$ (denoted by node $j$).
(3) Keep the current matching and temporarily remove node $i$ with all its links. Check if there is an augmenting path \cite{Karp-IEEE-81,Hopcroft-SIAM-73} that starts from node $j$, ends at an unmatched node and alternates between unmatched and matched links on the path.
(4) If no augmenting path is found, node $i$ needs to be always matched, therefore it is redundant. Otherwise node $i$ is replaceable and hence it is intermittent.
(5) Add back the removed node $i$ and repeat step (2) until all nodes in set $M$ is tested. 

The number of computational steps needed to find one maximum matching is $\mathcal O(N^{0.5}L)$. Each matched node requires a breadth first search ($\mathcal O(L)$ time) for the augmenting path. The number of matched nodes is proportional to $N$. Therefore the complexity of this algorithm is $\mathcal O(NL)$.

{\bf Calculating $n_\text{r}$ analytically.} 
Denote with $G$ a bipartite graph and $G^{\prime} = G\backslash i$ the subgraph of $G$ obtained by removing node $i$ and all its links (Supplementary Fig. S6). We proved a theorem that node $i$ is not always matched in $G$ if and only if all its neighboring nodes are always matched in the subgraph $G^{\prime}$ (Supplementary Note 6).
Based on this theorem, the probability that a node in the $in$ set with degree $k$ is not always matched is $p_{k,\text{in}} = (\theta_{\text{out}})^{k}$ where $\theta_{\text{out}}$ is the probability that a neighboring node is always matched in the $out$ set of the subgraph $G^{\prime}$. When averaging over degree distributions in infinite networks, the fraction of redundant nodes, which are always matched nodes in the $in$ set, is
\begin{eqnarray}
n_r &=& 1- \sum_{k=0}^{\infty} P_{\text{in}}(k) \times p_{k,\text{in}} \nonumber \\
&=&1- \sum_{k=0}^{\infty} P_{\text{in}}(k)(\theta_{\text{out}})^{k} = 1-G_{\text{in}}(\theta_{\text{out}}) \label{eq:method_nr_analytical},
\end{eqnarray}
which gives equation (\ref{eq:nr_analytical}).

In the construction of subgraph $G^{\prime}$, a neighbor of node $i$ is accessed through a randomly chosen link. This implies that the degree distribution associated with the neighboring node is the excess degree distribution, {\it i.e.} the degree distribution for a node at the end of a randomly chosen link. Furthermore, in $G^{\prime}$ this randomly chosen link is excluded, meaning that the degree of the neighboring nodes will be less by 1. Therefore $\theta_{\text{out}}$ satisfies the equation
\begin{equation}
1-\theta_{\text{out}} = \sum_{k=1}^{\infty}Q_{\text{out}}(k)(\theta_{\text{in}})^{k-1} = H_{\text{out}}(1-\theta_{\text{in}}).\label{eq:beta_1}
\end{equation}
By substituting $in$ for $out$, we can build a similar equation
\begin{equation}
1-\theta_{\text{in}} = \sum_{k=1}^{\infty}Q_{\text{in}}(k)(\theta_{\text{out}})^{k-1} = H_{\text{in}}(\theta_{\text{out}}).\label{eq:beta_2}
\end{equation}
Combining equations (\ref{eq:beta_1}) and (\ref{eq:beta_2}), we obtain equation (\ref{eq:beta_equation}).

{\bf Networks analyzed.} The model networks in this paper are Erd\H{o}s-R\'enyi network \cite{Erdos-PMIHAS-60} and scale-free network~\cite{Barabasi-Science-99,Caldarelli-Book-07} generated via the static model \cite{Goh-PRL-01}. The real networks we explored are described in Supplementary Table S1 and S2.

{\bf Acknowledgements}

We thank C. Song and D. Wang for discussions. We especially thank M. Martino for the thumbnail image. This work was supported by the Network Science Collaborative Technology Alliance sponsored by the US Army Research Laboratory under Agreement Number W911NF-09-2-0053; the Defense Advanced Research Projects Agency under Agreement Number 11645021; the Defense Threat Reduction Agency award WMD BRBAA07-J-2-0035; and the generous support of Lockheed Martin. M.~P\'{o}sfai has received funding from the European Union Seventh Framework Programme (FP7/2007-2013) under grant agreement No. 270833. E. Cs\'oka has received funding from European Research Council under grant agreement No. 306493, MTA Renyi "Lendulet" Groups and Graphs Research Group, ERC Advanced Research Grant under grant No. 227701 and KTIA-OTKA under grant No. 77780.

{\bf Author contributions}

T.J., Y.-Y.L., J.-J.S. and A.-L.B. designed the research. T.J. performed numerical simulation and analyzed the empirical data. T.J., Y.-Y.L., E.C. and M.P. did the analytical calculations. T.J. and A.-L.B. prepared the initial draft of the manuscript. All authors contributed comments on the results and revisions to the final version.

{\bf Competing financial interests:} The authors declare no competing financial interests.

\newpage

\noindent

\newpage
\noindent

\newpage
\noindent


\end{document}